# Hot electron cooling in three-dimensional Dirac fermion systems at low temperature: effect of screening

K. S. Bhargavi[1] and S. S. Kubakaddi[2*]

[1] Department of Physics, Siddaganga Institute of Technology, Tumkur-572013, Karnataka, India
[2] Department of Physics, Karnatak University, Dharwad- 580003, Karnataka, India

**Abstract:** Hot electron cooling rate $P$, due to acoustic phonons, is investigated in three-dimensional Dirac fermion systems at low temperature taking account of screening of electron-acoustic phonon interaction. $P$ is studied as a function of electron temperature $T_e$ and electron concentration $n_e$. Screening is found to suppress $P$ very significantly for about $T_e < 0.5$ K and its effect reduces considerably for about $T_e > 1$K in $Cd_3As_2$. In Bloch-Gruniesen (BG) regime, for screened (unscreened) case $T_e$ dependence is $P \sim T_e^9$ ($T_e^5$) and $n_e$ dependence gives $P \sim n_e^{-5/3}$ ($n_e^{-1/3}$). The $T_e$ dependence is characteristics of 3D phonons and $n_e$ dependence is characteristics of 3D Dirac fermions. In BG regime, screening effect is found to enhance for larger $n_e$. Screening is found to reduce the range of validity of BG regime temperature. Plot of $P/T_e^4$ vs $T_e$ shows a maximum at temperature $T_{em}$ which shifts to higher values for larger $n_e$. Interesting observation is that maximum of $P/T_e^4$ is nearly same for different $n_e$ and $T_{em}/n_e^{1/3}$ appears to be nearly constant. More importantly, we propose, the $n_e$ dependent measurements of $P$ would provide a clearer signature to identify 3D Dirac semimetal phase.

## I. INTRODUCTION

During the last couple of years enormous interest is focused on three-dimensional Dirac semimetals (3DDS) phase, the 3D analogue of graphene, as they are predicted to host many exciting physical phenomena. 3D Dirac semimetals have linear gapless massless electron and hole bands meeting at Dirac point. Predictions of 3D Dirac semimetals [1,2] are realized experimentally in $Na_3Bi$ [3] and $Cd_3As_2$ [4-6] and these are the only two widely and extensively studied. More recently, Chen et al [7] fabricated a high quality 3DDS $Cd_3As_2$ microbelts by a chemical vapour deposition method, with ultrahigh electron mobility $2\times10^4 cm^2/V$ s at 300 K. This is expected to open up a new avenue for fabrication of scalable $Cd_3As_2$ materials towards exciting electronic applications of 3D Dirac semimetals. Very recently, measured ultrahigh mobility $9\times10^6$ cm$^2$ V$^{-1}$ s$^{-1}$ at 5 K is reported in Dirac semimetal $Cd_3As_2$ which is attributed to a remarkable mechanism that strongly suppresses backscattering of the carriers [8]. There exists theoretical work discussing the transport properties of 3D Weyl and Dirac semimetals [9,10]. Using the semi-classical Boltzmann transport theory in the relaxation time approximation, an investigation of screening effects on transport relaxation time of 3D Dirac semimetals has been made, considering the momentum relaxation processes due to scattering by short-range and long-range disorder, and acoustic phonons [10]. Scattering by acoustic phonons via deformation potential coupling is shown to be dominant one and limits the mobility at higher temperature. Strength of this mechanism is determined by electron-phonon deformation potential coupling constant $D$, a key parameter rarely addressed in bulk $Cd_3As_2$ semiconductor. There is only one paper making an attempt to determine $D$ (10-30 eV) by fitting the theory of thermopower via mobility to the experimental data in bulk $Cd_3As_2$ [11].

---

*sskubakaddi@gmail.com

Study of transport in high electric field in 3DDS is yet to begin. In high electric field, electrons gain energy and thermalize rapidly among themselves leading to the establishment of 'hot electron' temperature $T_e$ which is greater than the lattice temperature $T$. In the steady state, these electrons lose their energy by emission of phonons. This hot electron energy relaxation /cooling takes place with acoustic and optical phonons as the dissipative channels. The electron heating effect has become extremely important for the device operation in the high field region. It affects the thermal dissipation and heat management in electronic devices. Electron cooling rate studies provide insight into the thermal link between electrons and phonons. Electron heating by photons finds potential applications in bolometry and calorimetry.

Hot electron relaxation, at low temperature, depends purely on electron-acoustic phonon interaction and is used as tool to determine the electron-phonon coupling strength. Hot electron cooling is extensively studied in bulk semiconductors [12-13], conventional two-dimensional electron gas (2DEG) of low-dimensional semiconductor heterostructures [14-21], graphene [22-35] and monolayer $MoS_2$[36]. In bulk semiconductors electron gas is three-dimensional (3DEG) and in the latter three systems electron gas is 2DEG. In graphene, experimental investigations of hot electron energy relaxation as function of electron temperature and carrier concentration [29, 30, 35] are used to determine the electron-acoustic phonon deformation potential coupling constant $D$ by comparing the experimental observations with the theoretical predictions [22,34]. Similarly, study of this property is used to determine $D$ in GaAs/GaAlAs [15,16] and Si/SiGe [20] heterojunctions (HJs). Very recently, hot electron cooling in Weyl and Dirac 3D semimetal, a key issue with the possible applications in electronic ultrafast and high field devices based on this material, by exchange of energy with acoustic phonons has been theoretically explored including the effect of disorder [37]. This study is without taking account of effect of screening of electron-acoustic phonon interaction. There is also a recent experimental report, by ultrafast transient measurements, on electron energy relaxation in $Cd_3As_2$ at higher temperature [38]. These observations suggest the cooling to proceed first through rapid emission of optical phonons, then through slower emission of acoustic phonons.

Effect of screening of electron-acoustic phonon interaction is expected to change the magnitude and temperature dependence of cooling very significantly, particularly at low temperatures as shown in semiconductor heterostructures [16,18]. We study, theoretically, hot electron cooling taking account of screening in 3D Dirac fermions, in absence of disorder, at low temperature. Its dependence on temperature and electron concentration is investigated and power laws are given in very low temperature regime both for screened and unscreened electron-acoustic phonon coupling. From these studies we make a significant proposal that these power laws with regard to electron concentration dependence can be used to identify the 3D Dirac phase of the material. Electron scattering by optical phonons in 3D Dirac materials is likely to be less significant for temperatures up to a few hundred Kelvin [37] and it is ignored in the present work.

## II. THEORETICAL FORMALISM OF COOLING POWER

In a 3D Dirac semimetal the low-energy linear electronic Dirac band dispersion $E_\mathbf{k} = \pm s\hbar v_f |\mathbf{k}|$, where $s=\pm 1$ (with +/- signs denoting electron/hole bands), $v_f$ is the Fermi velocity, $\mathbf{k}$ is the 3D wave vector of the electron. The corresponding electron velocity $\mathbf{v_k} = (1/\hbar)\nabla_\mathbf{k} E_\mathbf{k}$ and the density of states $D(E_\mathbf{k}) = gE_\mathbf{k}^2/[2\pi^2(\hbar v_f)^3]$, where $g$ is the degeneracy factor. The conduction and valence band meet at the Dirac point and make the system a gapless semiconductor or a semimetal with the zero effective mass of the carriers.

We work in the electron temperature model in which electrons are assumed to obey Fermi-Dirac statistics with temperature $T_e$ which is given by $f(E_\mathbf{k}) = [\exp[(E_\mathbf{k}-E_f)/k_B T_e]+1]^{-1}$. Following the technique given in [12,15,22], the average electron energy loss rate $<dE_\mathbf{k}/dt> = P$ can be obtained by calculating the total energy gained by the phonons from the electrons and dividing by the total number of electrons $N_e$. It is given by $P = (1/N_e)\sum_\mathbf{q} \hbar\omega_\mathbf{q} (dN_\mathbf{q}/dt)$, where $dN_\mathbf{q}/dt$ is the rate of change of phonon distribution $N_\mathbf{q}$ due to electron-phonon coupling. We assume 3D longitudinal acoustic phonons of energy $\hbar\omega_\mathbf{q}$ and wave vector $\mathbf{q}$ to interact with 3D Dirac electrons via deformation potential coupling [10]. The corresponding matrix element is given by $|C(q)|^2 = (D^2\hbar q/2V\rho v_s)(1+\cos\theta)/2$, $V$ is the volume of the sample, $\rho$ is the mass density, $v_s$ is the velocity of the phonon and $\theta$ is the angle between the initial $\mathbf{k}$ and final $\mathbf{k}'$ state vectors of the electron.

The rate of change of phonon distribution due to electron-acoustic phonon interaction is given by

$$\left(\frac{dN_\mathbf{q}}{dt}\right) = \frac{2\pi}{\hbar} g \sum_\mathbf{k} \frac{|C(q)|^2}{\varepsilon^2(q)} \{(N_\mathbf{q}+1)f(E_\mathbf{k}+\hbar\omega_\mathbf{q})[1-f(E_\mathbf{k})]$$
$$- N_\mathbf{q} f(E_\mathbf{k})[1-f(E_\mathbf{k}+\hbar\omega_\mathbf{q})]\}\delta(E_\mathbf{k+q}-E_\mathbf{k}-\hbar\omega_\mathbf{q})\delta_{\mathbf{k'},\mathbf{k+q}} \quad (1)$$

where $\varepsilon(q)$ is the screening function. In the equation for $P$, due to Kronecker delta function, the summation over $\mathbf{q}$ is replaced by the summation over $\mathbf{k}'$, which inturn is replaced by the integral

$$\sum_{\mathbf{k'}} \to \frac{V}{8\pi^3} \int_0^\infty k'^2 dk' \int_0^\pi \sin\theta d\theta \int_0^{2\pi} d\phi = \frac{V}{8\pi^3(\hbar v_f)^3} \int_0^\infty dE_{\mathbf{k'}} E_{\mathbf{k'}}^2 \int_0^\pi \sin\theta d\theta \int_0^{2\pi} d\phi$$
$$. \quad (2)$$

Integration with respect to $E_{\mathbf{k'}}$ is carried out using Dirac delta function. Converting the $\theta$ integration into $q$ integration, in the quasielastic approximation [10] (which gives $|Sin(\theta/2)| = q/2k$), we obtain

$$P = \frac{g}{2\pi n_e \hbar(\hbar v_f)^3} \sum_\mathbf{k} \int_0^{(\hbar\omega_q)_{max}} dq \left(\frac{q}{2k}\right)^2 (E_\mathbf{k}+\hbar\omega_\mathbf{q})^2 \frac{|C(q)|^2}{\varepsilon^2(q)} \hbar\omega_\mathbf{q},$$
$$\times [N_\mathbf{q}(T_e)-N_\mathbf{q}(T)][f(E_\mathbf{k})-f(E_\mathbf{k}+\hbar\omega_\mathbf{q})], \quad (3)$$

where $n_e = N_e/V$ electron concentration, $(\hbar\omega_\mathbf{q})_{max} = (\hbar\omega_\mathbf{q})_{q=2k} = 2v_s E_\mathbf{k}/v_f$ and $N_\mathbf{q}(T) = [\exp(\hbar\omega_\mathbf{q}/k_B T)-1]^{-1}$ is the Bose distribution at temperature $T$. In obtaining Eq.(3), the identity $(N_\mathbf{q}+1)f(E_\mathbf{k}+\hbar\omega_\mathbf{q})[1-f(E_\mathbf{k})]-N_\mathbf{q} f(E_\mathbf{k})[1-f(E_\mathbf{k}+\hbar\omega_\mathbf{q})] = [N_\mathbf{q}(T_e)-N_\mathbf{q}(T)][f(E_\mathbf{k})-f(E_\mathbf{k}+\hbar\omega_\mathbf{q})]$ has been used. Now, the summation over $\mathbf{k}$ is converted to integration over energy $E_\mathbf{k}$, which gives

$$\sum_\mathbf{k} \to \frac{V}{8\pi^3(\hbar v_f)^3} \int_0^\infty dE_\mathbf{k} E_\mathbf{k}^2 \int_0^\pi \sin\varphi d\varphi \int_0^{2\pi} d\psi = \frac{V}{2\pi^2(\hbar v_f)^3} \int_0^\infty dE_\mathbf{k} E_\mathbf{k}^2. \quad (4)$$

Then, using the matrix element $|C(q)|^2 = (D^2\hbar q/2V\rho v_s)[1-(q/2k)^2]$, we obtain

$$P = \frac{gD^2}{8\pi^3 \rho \hbar^7 n_e v_f^4 v_s^4 \rho \pi^2} \int_0^\infty dE_\mathbf{k} \int_0^{(\hbar\omega_q)_{max}} d(\hbar\omega_\mathbf{q})(\hbar\omega_\mathbf{q})^3 \frac{(E_\mathbf{k}+\hbar\omega_\mathbf{q})^2}{\varepsilon^2(q)}$$
$$\times g(q,k)[N_\mathbf{q}(T_e)-N_\mathbf{q}(T)][f(E_\mathbf{k})-f(E_\mathbf{k}+\hbar\omega_\mathbf{q})], \quad (5)$$

where $g(q,k) = [1-(\hbar\omega_\mathbf{q}/E_\mathbf{k})^2 (v_f/2v_s)^2]$. The temperature independent screening function $\varepsilon(q) = 1+(q_{TF}/q)^2$, where $q_{TF} = (4\pi e^2 D(E_f)/\varepsilon_s)^{1/2}$ [10] is the Thomas-Fermi screening wave vector and $\varepsilon_s$ is the static dielectric constant of the material.

At very low temperatures, the Bloch-Gruneisen (BG) regime ($q<<2k_f$ and $\hbar\omega_\mathbf{q} \approx k_B T$) is characterised by temperature $T_{BG} = 2v_s k_f/k_B$, where $k_f = (6\pi^2 n_e/g)^{1/3}$ is the Fermi wave vector. For $T<< T_{BG}$, we approximate $f(E_\mathbf{k})-f(E_\mathbf{k}+\hbar\omega_\mathbf{q}) \approx \hbar\omega_\mathbf{q} \delta(E_\mathbf{k}-E_f)$, $(E_\mathbf{k}+\hbar\omega_\mathbf{q}) \approx E_\mathbf{k}$ and $g(q,k) \approx 1$. Screening function is also approximated to be $\varepsilon(q) \approx (q_{TF}/q)^2$. Then, expressing $P = F(T_e)-F(T)$ [16,18], we have

$$F(T_e) = F^0 4!\zeta(5) T_e^\delta \quad \text{with} \quad \delta=5 \quad (6a)$$

for unscreened electron-phonon coupling and

$$F(T_e) = F^0 S^2 8!\zeta(9) T_e^\delta \quad \text{with} \quad \delta=9 \quad (6b)$$

for screened electron-phonon coupling. Here

$$F^0 = [gD^2 E_f^2 k_B^5]/[8\pi^3 \rho \hbar^7 n_e v_s^4 v_f^4], \quad S=(k_B/q_{TF}\hbar v_s)^2. \quad (6c)$$

It is found that

$$F(T_e) \sim (n_e)^{-p}, \quad (7)$$

where $p=5/3$ (1/3) with (without) screening.

We express $T_{BG}= 37.5 v_s n_e^{1/3}$ K, with $v_s$ ($n_e$) taken in units of $10^6$ cm/s ($10^{18}$ cm$^{-3}$). In Cd$_3$As$_2$ Dirac 3D semimetal $T_{BG}= 8.627$ K for $v_s= 2.3 \times 10^5$ cm/s and $n_e= 1 \times 10^{18}$ cm$^{-3}$.

## III.  RESULTS AND DISCUSSION

In the following, hot electron cooling rate is numerically investigated as a function of electron temperature $T_e$ (= 0.1-50 K) and electron concentration $n_e$ = (1- 30)x$10^{17}$ cm$^{-3}$ taking lattice temperature $T$=0 K. The $n_e$ range is restricted for which $E_f$ will be less than the band inversion energy scale of about 250 meV [2,37] as the Dirac description of Cd$_3$As$_2$ is only applicable below this energy scale. These evaluations are made in 3D Dirac semimetal Cd$_3$As$_2$ using the reasonably known material parameters : $D$= 20 eV [11], $\rho$ =7.0 gm/cm$^3$, $v_s$= 2.3x$10^5$ cm/s, $v_f$= 1x$10^8$cm/s [37], $\varepsilon_s$=36 [39] and $g$=4 [37]. To bring out the effect of screening, $P$ calculations are presented with and without (i.e. $\varepsilon(q)$=1) the screening of electron-acoustic phonon coupling. We use temperature independent screening function, in the present work for illustration. It is reasonable, in the temperature range of interest considered, in view of the demonstration of  electrical conductivity calculations showing no effect of temperature dependence of screening for $T_e \leq 0.1 T_f$, where $T_f$ ( = $E_f/k_B$= 2320 K for $E_f$ =200 meV) is the Fermi temperature [10].

In Fig. 1, $P$ is plotted as a function of $T_e$ for $n_e$=1x$10^{18}$cm$^{-3}$. The curves are shown for screened and unscreened interaction along with the respective BG regime power laws. In the very low $T_e$ region, for both screened and unscreened coupling case, $P$ increases very rapidly. Then, at higher $T_e$ this dependence becomes weaker and reaches nearly a linear behaviour. $P$, with screening, is found to increase more rapidly with the increasing temperature than unscreened $P$ does. This increase of $P$, with screening, is more rapid in 3DDS as compared to semiconductor HJs [16] and Si-MOSFETs [18] due to stronger $q$ dependence of screening function in the former case. The rapid increase, at lower $T_e$, may be attributed to the increasing number of phonons as their wave vector $q \approx k_B T_e/\hbar v_s$ increases linearly with $T_e$. Screening is found to suppress $P$ very significantly at very low $T_e$. For eg . at $T_e$= 0.1 K, $P$ is reduced by about two orders of magnitude and at 1 K the reduction is by about 1.5times. For still higher temperatures the effect of screening tends to be still smaller in Cd$_3$As$_2$. Whereas, screening is found to be still important in GaAs HJs [16] and Si-MOSFETs [18] at higher temperatures. In this context it is to be noted that in Cd$_3$As$_2$ the dielectric constant $\varepsilon_s$=36 is relatively large, which is nearly three times that in GaAs HJs and Si-MOSFETs.

In the Bloch-Gruneisen regime $P \sim T_e^\delta$ with $\delta$= 9(5) for screened (unscreened) electron-phonon coupling. Acoustic-phonon limited resistivity in BG regime shows the same temperature dependence [10]. The value of $\delta$= 5 is characteristics of 3D phonons, as found in GaAs/GaAlAs HJs [16] and Si-MOSFETs [18], for the unscreened deformation potential coupling. This is in contrast with $\delta$= 4 in graphene [12,34] and monolayer MoS$_2$ [36] in which phonons are  2D in nature. For the screened case, the value of $\delta$= 9 in Cd$_3$AS$_2$, in contrast with $\delta$= 7 in GaAs/GaAlAs [16] and Si/SiGe [21] HJs and Si-MOSFETs [18] and $\delta$ = 6 in monolayer MoS$_2$ [36]. This large enhancement of power in Cd$_3$As$_2$ is due to screening function, in the limit as $q\rightarrow 0$, $\varepsilon^2(q) \sim q^{-4}$, with $q \sim T$. In the same limit $\varepsilon^2(q) \sim q^{-2}$ in low-dimensional semiconductor heterostructures [16,18]. We see from Fig. 1 that, for $n_e$=1x$10^{18}$ cm$^{-3}$, for the unscreened case $P \sim T_e^5$ law is valid for $T_e$ up to about 0.4 K, where as for screened case $P \sim T_e^9$ is valid only for about $T_e$ below  0.1 K. This indicates that screening reduces the temperature range of BG regime considerably. $P$  is found to be very much sensitive to phonon group velocity. In BG regime for  screened (unscreened) case $P \sim v_s^{-8}$ ($v_s^{-4}$) as compared to $v_s^{-6}$ ($v_s^{-4}$) in conventional 2DEG [16,18], $v_s^{-3}$ (unscreened) in graphene [22] and $v_s^{-5}$ ($v_s^{-3}$) for screened (unscreened) coupling in monolayer MoS$_2$[36].

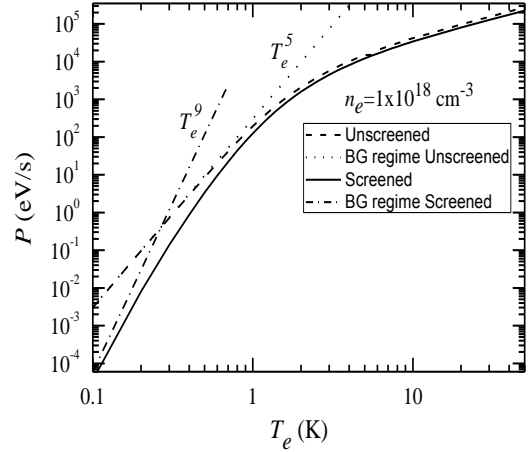

**Figure 1:** Electron cooling rate $P$ vs electron temperature $T_e$.

It is to be noted that the analytical results of power loss study  in Ref. [37] don't give any power law either for $T_e$ or $n_e$ dependence. Their results are for $T_e$ above BG regime. However,  short-range disorder assisted power loss shows $T_e^4$ with drastic enhancement, as there is no restriction on phonon momentum [37].

Comparison of magnitude of $P$ in 3D Dirac fermions in Cd$_3$AS$_2$ for $n_e$=1x$10^{18}$ cm$^{-3}$ with that in monolayer graphene (i.e.2D Dirac fermions) for $n_s$=1x$10^{12}$ cm$^{-2}$ ($n_s$ being surface electron concentration), at 1 K,  shows that $P$ in the latter case is about 50 times smaller. This is significant, although in 3D Dirac fermions $P$ is due to screened coupling where as in graphene coupling is taken to be unscreened [22].This difference may be attributed mainly to the larger $v_s$ in graphene which is about an order of magnitude larger than that in Cd$_3$As$_2$. Broadly, we observe,  values of  $P$ (about $10^3$ eV/s at 1 K and $10^6$ eV/s at 10 K) in monolayer MoS$_2$ [36] are much larger than the respective $P$ values of 2DEG  in monolayer graphene [22], GaAs/GaAlAs HJs [16] Si-MOSFETs[18] and 3DEG of Dirac fermions in  Cd$_3$As$_2$ (present work). Possible reasons are, in monolayer MoS$_2$, (i) dominant $P$ is due to unscreened transverse acoustic phonon coupling and (ii) velocity of these phonons is relatively smaller.

In Figs. 2(a) and 2(b), $P$ vs $T_e$ is shown, respectively, for screened and unscreened electron-phonon coupling, for $n_e$= 1,5,10x$10^{17}$ cm$^{-3}$. In the low $T_e$ region, $P$ is smaller for larger $n_e$ and with the increasing $T_e$ the cross over is taking place reversing this trend. This may be attributed to power law being valid for larger $T_e$ range for larger $n_e$, before the smooth transition is made to the behaviour in the higher $T_e$ region. In the low $T_e$ region the difference of $P$ for different $n_e$ is smaller for unscreened coupling compared to screened case due to weaker $n_e$ dependence in the former. In the higher $T_e$ region the difference in $P$ in the screened and unscreened values due to different $n_e$ is nearly same as screening effect becomes weaker.

region indicate the onset of higher power dependence. Each curve is exhibiting a maximum at a particular temperature $T_{em}$ which is shifting to higher $T_e$ values for larger $n_e$. Interestingly, at $T_{em}$, the magnitude of $P/T_e^4$ for all the three $n_e$ is found to be nearly same with $T_{em}/n_e^{1/3}$ nearly a constant value about 1x$10^{-6}$ K-cm. For a given $n_e$, the maximum of $P/T_e^4$, for screened and unscreened case, appears to occur nearly at the same $T_{em}$. Similar observations are made in a recent study of phonon-drag thermopower in this material [40]. We also note that matrix element $g(q, k)$, arising from the spinor wave function, reduces the value at the maximum by nearly two times. This effect is $T_e$ dependent as found in bilayer [34] graphene.

We make the comparison of $T_{BG}$ value in Cd$_3$As$_2$ with those in graphene and GaAs heterojunctions. In this connection, it is important to note that $T_{BG}$ depends only upon electron concentration and phonon velocity $v_s$. In Cd$_3$As$_2$ $T_{BG}$= 8.627 K for $n_e$=1x$10^{18}$ cm$^{-3}$ (i.e. for $k_f$ =2.455x$10^6$ cm$^{-1}$ and $E_f$ =160 meV) as compared to $T_{BG}$ = 54.153 K for 2D electron concentration $n_s$=1x$10^{12}$ cm$^{-2}$ (i.e. for $k_f$ = 1.722x$10^6$ cm$^{-1}$ and $E_f$ = 117 meV) in

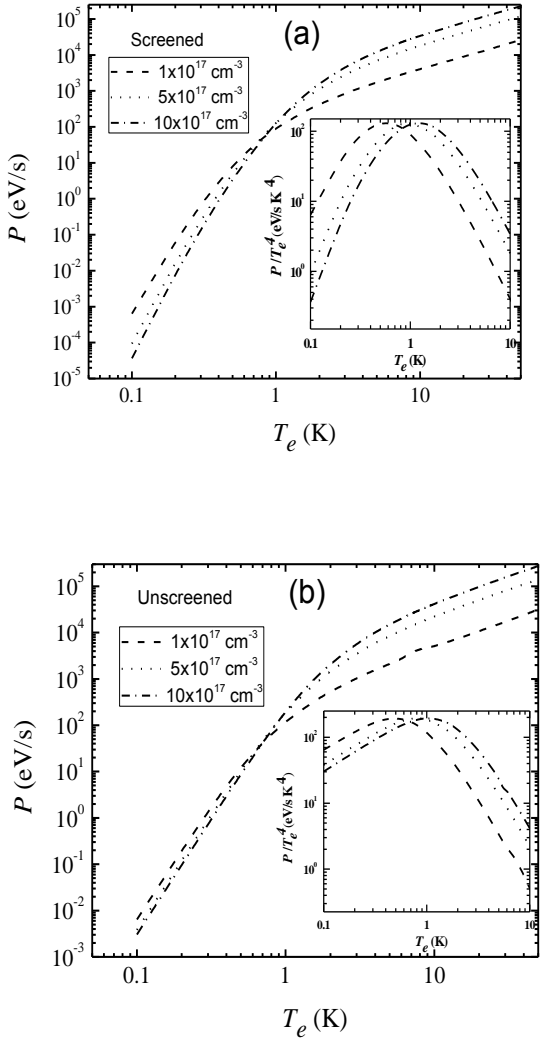

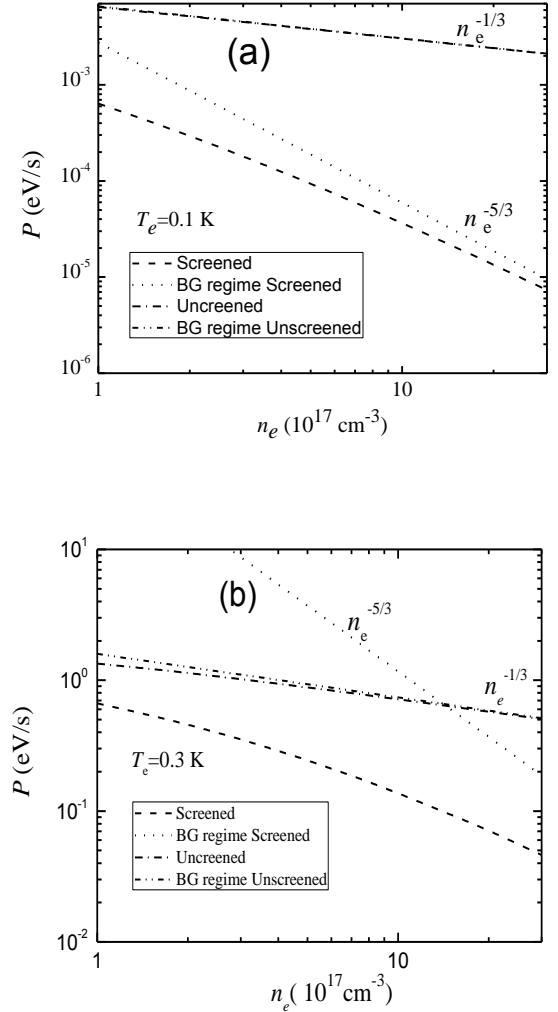

**Figure 2:** Electron cooling rate $P$ vs electron temperature $T_e$ for different $n_e$. (a) Screened (b)Unscreened. Inset : $P/T_e^4$ vs $T_e$.

In the inset of Fig. 2, we show $P/T_e^4$ vs $T_e$ for $n_e$ = 1,5,10x$10^{17}$ cm$^{-3}$, in which major $T_e^4$ dependence is removed. Deviations from this behaviour in low $T_e$ monolayer graphene [22]. This difference may be attributed to the large $v_s$ ( ~ $2 \times 10^6$ cm/s) in graphene, an order of magnitude greater than that in $Cd_3As_2$, although $k_f$ in $Cd_3As_2$ is nearly 1.5 times that in monolayer graphene. In GaAs HJs, for $n_s = 1 \times 10^{12}$ cm$^{-2}$ ($k_f = 2.5 \times 10^6$ cm$^{-1}$, $E_f = 35.51$ meV), $T_{BG} = 19.67$ K for LA phonons with $v_s = 5.14 \times 10^5$ cm/s. Although $k_f$ in $Cd_3As_2$ and GaAs HJs is nearly same, the difference in $T_{BG}$ is attributed to the difference in $v_s$.

**Figure 3.** Electron cooling rate $P$ vs electron concentration $n_e$. (a) $T_e$= 0.1 K (b) $T_e$ = 0.3 K.

In Fig. 3 (a), we present screened and unscreened $P$ as a function of $n_e$ at $T_e = 0.1$ K along with the respective BG regime curves. The dependence of $P$ on $n_e$ is stronger at lower $T_e$ and becomes weaker at higher $T_e$. $P$ is found to decrease with the increasing $n_e$, more rapidly in screened coupling case than unscreened $P$ does. This is attributed to the additional $n_e$ dependence coming from $q_{TF}$, in the screening function, which is proportional to the square root of the density of states and hence $E_f$. The $P$ curve corresponding to unscreened case coincides with that of BG regime for the range of $n_e$ considered, indicating the perfect validity of BG regime in this case at $T_e = 0.1$ K . The curve corresponding to screened case tend to differ with BG regime curve less (more) in higher (lower) $n_e$ range. The screening effect is found to enhance with the increasing $n_e$. For eg, at $T_e = 0.1$ K, $P$ with screening is about two (three) orders of magnitude smaller than the $P$ without screening for $n_e = 1 \times 10^{17}(10 \times 10^{17})$ cm$^{-3}$. In BG regime, screened (unscreened) $P \sim n_e^{-5/3}(n_e^{-1/3})$ in contrast with $n_s^{-1/2}$ in monolayer graphene [22] and $n_s^{-3/2}$ in GaAs HJs [16], Si-MOSFETs [18], bilayer graphene [34] and monolayer $MoS_2$ [36]. This difference is attributed to different dimensionality of the electron gas, nature of dispersion and screening function dependence on carrier concentration in 2D and 3D systems.

In Fig. 3(b) $P$ vs $n_e$ is shown at $T_e = 0.3$ K. It is noticed that, $P$ due to unscreened coupling is still coinciding with that due to BG regime curve in the larger $n_e$ region and differing slightly in the lower $n_e$ region. Where as, there is large difference in the corresponding curves due to screening. This difference is found to be larger for smaller $n_e$.

We also like to point out that, in conventional 3D semiconductors, for which energy dispersion is quadratic, the screening function has the same form as given here for 3DDS, but $q_{TF} \sim n_e^{1/6}$ [41] as $D(E_f) \sim E_f^{1/2}$. In the BG regime this makes the change in $n_e$ dependence by $n_e^{2/3}$ in the denominator, as compared to the change by $n_e^{4/3}$ in 3DDS. However, $T_e$ dependence in 3D semiconductors will be changed by $T_e^4$ as in 3DDS. Hence, $n_e$ dependent measurements, rather than $T_e$ dependent, at very low $T_e$, may be used as tool to identify 3DDS phase as has been done in graphene [35].

Our predictions of $T_e$ and $n_e$ dependence of $P$, at low $T_e$, may be used as better tool for determination of precise value of $D$, in $Cd_3As_2$ Dirac semimetal, by comparing with the experimental results as done in graphene [29,30,35]. We feel there is need for more low $T_e$ experimental data of transport properties, that depend upon only electron-acoustic phonon interaction, to determine $D$ which may set ultimate upper limit on intrinsic mobility of $Cd_3As_2$ Dirac semimetal. Phonon-drag thermopower $S^g$ is another such transport property which purely depends upon electron-acoustic phonon coupling and been studied by us in this system very recently [40]. The expression for $F(T)$ and $S^g$ derived in Ref. [40] are based on the same basic assumptions and are related, at very low $T$, by $F(T) = -\xi S^g (v_s eT/\Lambda)$, $\xi$ is a numerical constant of the order of unity, $e$ is the charge of electron and $\Lambda$ is the phonon mean free path. This relation is same as found in silicon inversion layer [18], where phonons are considered to be 3D, and monolayer graphene [22].

## IV. CONCLUSIONS

In summary, hot electron cooling rate $P$ is investigated in 3D Dirac semimetals ($Cd_3As_2$), taking account of the effect of screening of electron-phonon interaction, in the temperature range $T_e = 0.1$-50 K and for electron concentration $n_e = (1$-$30) \times 10^{17}$ cm$^{-3}$ for lattice temperature $T=0$. Screening is found to suppress $P$ very significantly for about $T_e < 1$ K. $P$ is found to increase rapidly with the increasing $T_e$ for about $T_e < 1$ K and showing weaker dependence at higher $T_e$. Chiral nature of the electrons is found to reduce $P$, by a factor maximum of 2. At very low $T_e$, $P$ increases with decreasing $n_e$. Power laws $P \sim T_e^9$ ($T_e^5$) and $n_e^{-5/3}(n_e^{-1/3})$ are obtained in the Bloch-Gruneissen regime for screened (unscreened) case. Besides, the strong dependence on acoustic phonon velocity $v_s$ i. e. $P \sim v_s^{-8}$ ($v_s^{-4}$) is shown. The BG temperature is found to be very small compared to graphene and screening is found to push the strictly valid BG regime to still lower $T_e$. Our results obtained in the present calculations for $T=0$ also remain valid for $T \neq 0$, as long as $T << T_e$. We believe that the $n_e$ dependent measurements of hot electron relaxation will provide clearer signature of 3D Dirac semimetal phase. Also we point out that the theory developed in this work is applicable to Weyl semimetals.

_______________________________________________

1. Z. J. Wang, Y. Sun, X. Q. Chen, C. Franchini, G. Xu, H. M. Weng, X. Dai, and Z. Fang, Phys. Rev. B **85**, 195320 (2012).
2. Z. Wang, H. M. weng, Q. S. Wu, X. Dai, and Z. Fang, Phys. Rev. B **88**, 125427, (2013)
3. Z. K. Liu, B. Zhou, Y. Zhang, Z. J. Wang, H. M. Weng, D. Prabhakaran, S.-K. Mo, Z. X. Shen, Z. Fang, X. Dai, Z. Hussain, and Y. L. Chen, Science **343**, 864 (2014).
4. M. Neupane, S.-Y. Xu, R. Sankar, N. Alidoust, G. Bian, C. Liu, I. Belopolski, T.-R. Chang, H.-T. Jeng, H. Lin, A. Bansil, F. Chou, and M. Z. Hasan, Nat. Commun. **5**, 3786, (2014).
5. Z. K. Liu, J. Jiang, B. Zhou, Z. J. Wang, Y. Zhang, H. M. Weng, D. Prabhakaran, S-K. Mo, H. Peng, P. Dudin, T. Kim, M. Hoesch, Z. Fang, X. Dai, Z. X. Shen, D. L. Feng, Z. Hussain and Y. L. Chen, Nat. Mater. 13, 677 (2014).
6. S. Borisenko, Q. Gibson, D. Evtushinsky, V. Zabolotnyy, B. B¨uchner, and R. J. Cava, Phys. Rev. Lett. **113**, 027603 (2014).
7. Z.-G. Chen, C. Zhang, Y. Zou, E.Zhang, L. Yang, F. Xiu and J. Zou, *Nano Lett.,* **15**, 5830 (2015)
8. T. Liang, Q. Gibson, M. N. Ali, M. Liu, R. J. Cava and N. P. Ong, Nat. Mater. **14,** 280 (2015).
9. B. Skinner, Phys. Rev. B **90**, 060202(R) (2014).
10. S. Das Sarma, E. H. Hwang, and H. Min, Phys. Rev. B **91**, 035201 (2015).

11. J. P. Jay-Gerin, M. J. Aubin and L. G. Caron, Phys. Rev. B **18**, 4542 (1978).
12. E. M. Conwell, *High Field Transport in Semiconductors*, Academic Press, New York (1967).
13. B. R. Nag, *Electron Transport in Compound Semiconductors* (Springer-Verlag, Berlin, Heidelberg, New York, Vol. **11** (1980)).
14. J. Shah, A. Pinczuk, A. C. Gossard, and W. Wiegmann, Phys. Rev. Lett. 54, 2045 (1985).
15. S. J. Manion, M. Artaki, M. A. Emanuel, J. J. Coleman, and K. Hess, Phys. Rev. B 35, 9203 (1987).
16. Y. Ma, R. Fletcher, E. Zaremba, M. D'Iorio, C. T. Foxon, and J. J. Harris, Phys. Rev. B **43** 9033 (1991).
17. B. K. Ridley, Rep. Prog. Phys. 54, 169 (1991).
18. R. Fletcher, V. M. Pudalov, Y. Feng, M. Tsaousidou and P. N. Butcher, Phys. Rev. B **56** 12422 (1997).
19. S. S. Kubakaddi, Kasala Suresha and B. G. Mulimani, Semicond. Sci. Technol. 17, 557 (2002).
20. D. Lehmann, A. J. Kent, Cz. Jasiukiewicz, A. J. Cross, P. Hawker, and M. Henini, Phys. Rev. B 65, 085320 (2002).
21. S. S. Kubakaddi, V. S. Katti and D. Lehmann, J. Appl. Phys. **107**, 123716 (2010).
22. S. S. Kubakaddi, Phys. Rev. B **79** 075417 (2009).
23. W.-K. Tse and S. Das Sarma, Phys. Rev. B 79, 235406 (2009).
24. R. Bistritzer and A. H. MacDonald, Phys. Rev. Lett. 102, 206410 (2009).
25. J. K. Viljas and T. T. Heikkil• a, Phys. Rev. B 81, 245404 (2010).
26. A. C. Betz, F. Vialla, D. Brunel, C. Voisin, M. Picher, A. Cavanna, A. Madouri, G. Feve, J.-M. Berroir, B. Placais, et al Phys. Rev. Lett. 109, 056805 (2012).
27. J. C. W. Song, M. Y. Reizer, and L. S. Levitov, Phys. Rev. Lett. 109, 106602 (2012).
28. W. Chen and A. A. Clerk, Phys. Rev. B 86, 125443 (2012).
29. A. M. R. Baker, J. A. Alexander-Webber, T. Altebaeumer, and R. J. Nicholas, Phys. Rev. B 85, 115403 (2012).
30. A. M. R. Baker, J. A. Alexander-Webber, T. Altebaeumer, S. D. McMullan, T. J. B. M. Janssen, A. Tzalenchuk, S. Lara-Avila, S. Kubatkin, R. Yakimova, C.-T. Lin, L.-J Li, and R. J. Nicholas, Phys. Rev. B 87, 045414 (2013).
31. I. V. Borzenets, U. C. Coskun, H. T. Mebrahtu, Y. V. Bomze, A. I. Smirnov, and G. Finkelstein, Phys. Rev. Lett. 111, 027001 (2013).
32. M. W. Graham, S.-F. Shi, D. C. Ralph, J. Park, and P. L. McEuen, Nature Phys. 9, 103 (2013).
33. A. C. Betz, S. H. Jhang, E. Pallecchi, R. Ferreira, G. F_eve, J.-M. Berroir, and B. Placais, Nature Phys. 9, 109 (2013).
34. K. S. Bhargavi and S. S. Kubakaddi, Physica E 56, 123 (2014).
35. J. Huang, J. A. Alexander-Webber, T. J. B. M. Janssen, A. Tzalenchuk, T. Yager, S. Lara-Avila, S. Kubatkin, R. L. Myers-Ward, V. D. Wheeler, D. K. Gaskill and R. J. Nicholas, J. Phys.: Condens. Matter **27,** 164202 (2015).
36. K. Kaasbjerg, K. S. Bhargavi, and S. S. Kubakaddi, Phys. Rev. B **90**, 165436 (2014).
37. R. Lundgren and G. A. Fiete, Phys. Rev. B 92, 125139 (2015).
38. C. P. Weber, E. Arushanov, B. S. Berggren, T. Hosseini, N. Kouklin, and A. Nateprov Appl. Phys. Lett. **106**, 231904 (2015).
39. J. P. Jay-Gerin, M. J. Aubin and L. G. Caron, Solid State Commun., **21**, 771 (1977).
40. S. S. Kubakaddi, J. Phys.: Condens. Matter **27**, 455801 (2015).
41. B. K. Ridley, *Electrons and Phonons in Multilayer Semiconductors,* (Cambridge University Press, 1997) p. 255.